# Kepler-1656b's Extreme Eccentricity: Signature of a Gentle Giant

Isabel Angelo[1,2], Smadar Naoz[1,2], Erik Petigura[1], Mason MacDougall[1], Alexander P. Stephan[3], Howard Isaacson[4,5], and Andrew W. Howard[6]

[1] Department of Physics & Astronomy, University of California Los Angeles, Los Angeles, CA 90095, USA
[2] Mani L. Bhaumik Institute for Theoretical Physics, University of California, Los Angeles, Los Angeles, CA 90095, USA
[3] Department of Astronomy, The Ohio State University, Columbus, OH 43210, USA
[4] Department of Astronomy, University of California Berkeley, Berkeley, CA 94720, USA
[5] Centre for Astrophysics, University of Southern Queensland, Toowoomba, QLD, Australia
[6] Department of Astronomy, California Institute of Technology, Pasadena, CA 91125, USA

## Abstract

Highly eccentric orbits are one of the major surprises of exoplanets relative to the solar system and indicate rich and tumultuous dynamical histories. One system of particular interest is Kepler-1656, which hosts a sub-Jovian planet with an eccentricity of 0.8. Sufficiently eccentric orbits will shrink in the semimajor axis due to tidal dissipation of orbital energy during periastron passage. Here our goal was to assess whether Kepler-1656b is currently undergoing such high-eccentricity migration, and to further understand the system's origins and architecture. We confirm a second planet in the system with $M_c = 0.40 \pm 0.09\ M_{\rm jup}$ and $P_c = 1919 \pm 27$ days. We simulated the dynamical evolution of planet b in the presence of planet c and find a variety of possible outcomes for the system, such as tidal migration and engulfment. The system is consistent with an in situ dynamical origin of planet b followed by subsequent eccentric Kozai–Lidov perturbations that excite Kepler-1656b's eccentricity gently, i.e., without initiating tidal migration. Thus, despite its high eccentricity, we find no evidence that planet b is or has migrated through the high-eccentricity channel. Finally, we predict the outer orbit to be mutually inclined in a nearly perpendicular configuration with respect to the inner planet orbit based on the outcomes of our simulations and make observable predictions for the inner planet's spin–orbit angle. Our methodology can be applied to other eccentric or tidally locked planets to constrain their origins, orbital configurations, and properties of a potential companion.

*Unified Astronomy Thesaurus concepts:* Exoplanet astronomy (486); Exoplanet dynamics (490); Extrasolar gaseous giant planets (509); Exoplanet detection methods (489); Exoplanet evolution (491); Exoplanet migration (2205)

*Supporting material:* machine-readable table

## 1. Introduction

Despite their absence in our own solar system, highly eccentric planets have been discovered in a number of exoplanet systems. In contrast to the low-eccentricity orbits of planets in our own solar system, with mean eccentricity $\langle e \rangle = 0.06$, giant planets in radial velocity surveys have been found to span the full possible range of $0 \leqslant e < 1$ for bound orbits, with eccentricities as high as $e = 0.9$ in some cases (e.g., Masuda 2017; Blunt et al. 2019; Schlecker et al. 2020). This range of eccentricities points to a diverse variety of formation scenarios (see Winn & Fabrycky 2015 for a review).

Adding to this picture, a subset of these eccentric planets are giants orbiting much closer to their host star than the solar system Jovians. Roughly 1% of stars possess "Hot Jupiters" (giant planets with $P = 1$–10 days; see Udry & Santos 2007; Cumming et al. 2008; Howard et al. 2010a). This number is even larger for giant planets with $P < 200$ days (10.5%; see Wittenmyer et al. 2016). This class of close-orbiting giant planets must be corroborated by theories of planet formation, which originally predicted giant planets to form and accrete material further out beyond their host star's "ice line" (e.g., Boss 1997; Rubie et al. 2015; Morbidelli et al. 2015).

Eccentric orbits are a relic of planet formation processes and/or the influence of another body—either a planet, brown dwarf, or star—in the system. Because giant planets possess significant envelopes, they must form in early disk phases when some gas is still present ($t < 10\,Myr$; see Wyatt 2008). Since the presence of this gas tends to dampen eccentricity (e.g., Artymowicz et al. 1991; Ward 1988; Lee & Chiang 2016), some postnebular eccentricity excitation mechanism is required to explain their eccentric orbits. For example, a planet can be perturbed into an eccentric orbit via disk migration, planet–planet scattering, or an outer perturber (e.g., Rasio & Ford 1996; Fabrycky & Tremaine 2007; Wu et al. 2007; Naoz et al. 2011, 2012; Shara et al. 2016; Dawson & Johnson 2018). At this point, the planet may undergo high-eccentricity migration, where it is subject to strong tidal forces at closest approach that dissipate energy and shrink the planet's orbit over time.

In this paper, we focus on Kepler-1656b (a.k.a KOI-367, KIC-4815520), a warm, highly eccentric sub-Saturn orbiting a Sun-like host star (Brady et al. 2018). Kepler-1656b's orbital configuration makes it a strong candidate for high-eccentricity migration and is likely a relic of the aforementioned postnebular eccentricity excitation.[7] Moreover, sub-Saturns like Kepler-1656b lie in the sparsely populated regime of planets between the size of

---

[7] While it is possible that the planet opened a gap during the disk phase and reduced the eccentricity damping, even then the eccentricity is not readily increased to ∼0.8 values (Ward 1988; Goldreich & Sari 2003; Teyssandier & Ogilvie 2016, 2017; Ragusa et al. 2018), and the expectation is that a more massive planet than Kepler-1656b is needed to open a gap (depending on the disk's viscosity, e.g., Armitage 2007; Crida & Morbidelli 2007; Edgar et al. 2007; Duffell & MacFadyen 2013).

Neptune and Saturn, with densities spanning over an order of magnitude for a fixed size (Petigura et al. 2017a, 2018a). Planets of these sizes pose a challenge to standard core nucleated accretion theory, which predicts planets with similar-mass cores to have undergone runaway accretion and accumulate much larger gas envelopes than observed (Pollack et al. 1996). For example, a $10\,M_\oplus$ core at 5 au initiates runaway gas accretion in just $10^6$ yr. While there are a number of theories that explain the envelope fractions seen in some sub-Saturns (see, for example, Lee et al. 2014; Lee & Chiang 2015, 2016; Millholland et al. 2020), the detailed origin scenarios for these planets remain a mystery. Specifically, a tension exists between theories that sub-Saturns and giant planets more generally form in situ, or accrete material further out from their host star and migrate inwards later on (e.g., Batygin et al. 2016). This tension extends to Kepler-1656b, whose birth scenario (i.e., in situ versus inward migration) should corroborate its eccentric orbit and a high reported core mass fraction (see Millholland et al. 2020).

Here we use Kepler-1656b as a case study of combining observational data and dynamical analysis to constrain a system's detailed architecture and dynamical history. Originally, Brady et al. (2018) predicted an upper limit of $\dot{\gamma} < 1.4$ ms$^{-1}$ yr$^{-1}$ on the long-term stellar radial velocity (RV) trend, with observations disfavoring—but not ruling out—a planetary or stellar companion in the system. In this study, more recent RV observations suggest underlying periodic RV behavior with period $P \gg 100$ days. We identify a giant planetary companion to Kepler-1656b and find that perturbations from said companion are driving Kepler-1656b into its eccentric orbit gently, i.e., without inducing high-eccentricity migration. In doing so, we identify two classes of outer companions to sub-Saturns and giant planets based on their signature on the inner planet. The first class, "gentle companions," nudge the inner planet into an eccentric orbit in situ, without inducing migration. The second class, "strong companions," has a more dramatic effect on the inner planet, causing it to migrate inwards and settle in a close-in, tidally locked orbit. These classes of companions produce observable signatures on their inner companion's orbital configuration, and can potentially sculpt large-scale trends in exoplanet system architectures.

We begin by confirming the existence and planetary nature of the outer companion Kepler-1656c in Section 2. We then describe the physical setup and initial conditions we used to simulate the dynamical evolution of Kepler-1656b in the presence of this companion in Sections 3.1 and 3.2 respectively. In Section 4, we predict observable properties of the system's orbital configuration and discuss the dynamical origins of Kepler-1656b and eccentric and tidally locked planets more generally. Finally, in Section 5 we discuss how our proposed origin scenarios fit into our understanding of planet formation and the current exoplanet population at large.

## 2. Observational Signatures of a Planetary Companion

### 2.1. High-resolution Imaging

We searched for companions to Kepler-1656b in imaging data and did not find evidence for a stellar companion in any of our images. To start, we used the Gaia database to search for nearby, comoving companions to Kepler-1656. We found no

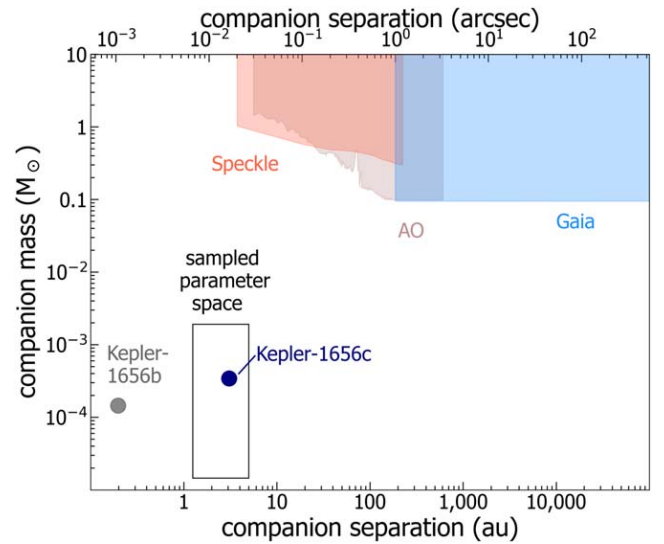

**Figure 1.** We consider regions of mass-separation parameter space in which a companion to Kepler-1656b can reside. Regions ruled out by Gaia (blue), AO (tan), and Speckle imaging (red) are shown. The explored region encapsulating Kepler-1656c is outlined, and both Kepler-1656b and Kepler-1656c are overplotted for reference.

**Table 1**
Radial Velocity and Activity Measurements

| Time | RV | $\sigma$(RV) | $S_{HK}$ |
| --- | --- | --- | --- |
| 2457942.954 | 13.892 | 1.922 | 0.152 |
| 2457943.009 | 0.213 | 1.966 | 0.149 |
| 2457944.956 | 16.476 | 1.819 | 0.154 |
| 2457945.004 | 11.981 | 2.275 | 0.148 |
| 2457945.994 | 17.432 | 2.066 | 0.149 |

**Note.** A portion of the data set is shown here. $S_{HK}$ errors are all set to 0.001, and RV uncertainties are photon limited.

(This table is available in its entirety in machine-readable form.)

targets within $10^5$ au of the host star. The Gaia DR2 catalog is complete to magnitudes as high as $G \sim 12$–18 (Gaia Collaboration et al. 2018), which conservatively rules out $M > 0.1$ $M_\odot$ companions beyond Gaia's 1″ detection limit. This is illustrated by the blue region in Figure 1.

We also searched for companions to Kepler-1656b using speckle and adaptive optics (AO) images of the host star. These images provide sensitivity to companion magnitudes as a function of radial distance from the host star. Following a similar process to that outlined in Furlan et al. (2017), we converted these sensitivities to corresponding companion masses using an interpolated magnitude-mass relationship from Pecaut & Mamajek (2013). Speckle images taken by the Gemini-N telescope with a 692 nm center wavelength filter effectively rule out companions residing in the red region of parameter space in Figure 1, and AO images from the Exoplanet Follow-up Observing Program (ExoFOP) archive[8] taken by the Keck II telescope $Ks$ filter rule out the tan region. As can be seen in Figure 1, imaging observations of Kepler-1656 effectively rule out stellar-mass companions. Thus, the dynamical picture we consider for the remainder of our analysis is a two-planet system in which Kepler-1656b is accompanied by a second planet in the system.

---
[8] https://exofop.ipac.caltech.edu/

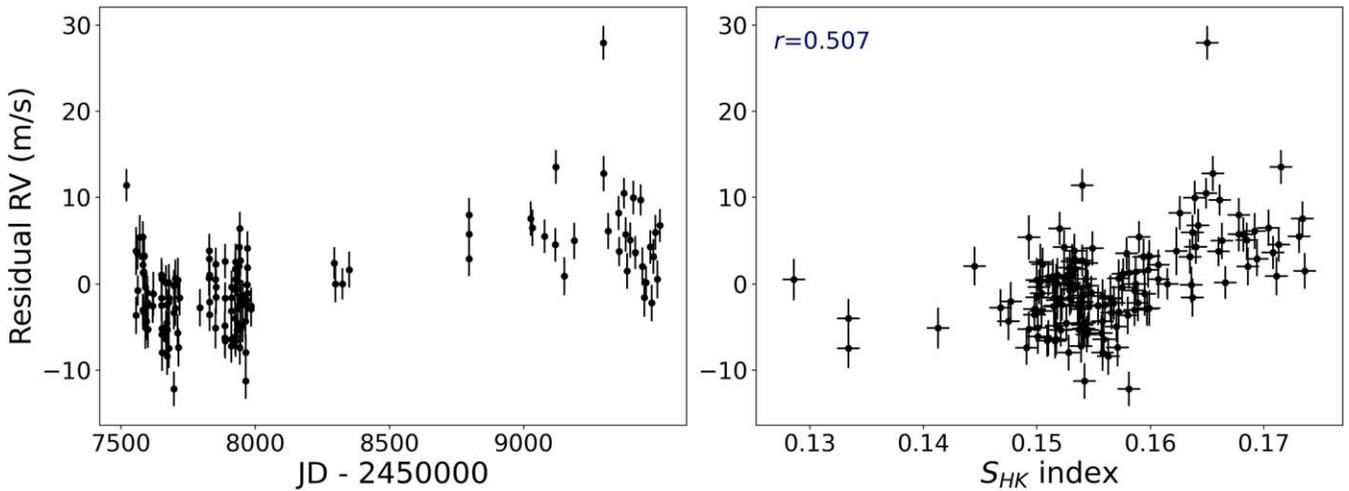

**Figure 2.** Left: residual RV time series of Kepler-1656 shows a long-term trend with $P \gg 100$ days. Right: residual RVs are plotted vs. $S_{HK}$. A Pearson $r$ statistic of 0.51 and median activity signal of $S_{HK} = 0.15$ indicate that stellar activity is not sufficient at explaining the variability in our RV time series.

### 2.2. Radial Velocities

Observations of Kepler-1656 were obtained through the California Planet Search (CPS; Howard et al. 2010b) using the High Resolution Echelle Spectrometer on the Keck I 10 m telescope (HIRES; Vogt et al. 1994). We collected 50 spectra between 2017 July 11 and 2020 October 24. These spectra were used along with 100 spectra reported in Brady et al. (2018) to compute RV time series data for Kepler-1656 using standard CPS methodology. For all observations, spectra were observed through an iodine cell mounted in front of the spectrometer for wavelength calibration, with an SNR of 110 per pixel at 500 nm. Our RVs are listed in Table 1.

Our RV observations reveal a long-term trend in addition to the short-period signal produced by Kepler-1656b that is suggestive of underlying periodic RV behavior with period $P \gg 100$ days. This behavior can be seen in the residual RVs after the inner planet signal has been subtracted (Figure 2, left panel).

We first considered the possibility that this observed trend was due to stellar activity. Variability in the magnetic field over the star's activity cycle can lead to apparent RV variability on multiyear timescales. In some cases, this variability can be mistaken for a planet signal (Haywood et al. 2014; Robertson et al. 2014). Using formalism outlined in Isaacson & Fischer (2010), we monitored stellar activity through Kepler-1656's $S_{HK}$ index, which traces the CaII H and K emission lines as an indicator of chromospheric activity (e.g., Wilson 1968; Duncan et al. 1991; Baliunas et al. 1995). Our results are shown in Table 1.

In the event that the observed trend is due to stellar activity, we should detect a strong correlation between the residual (i.e., inner planet subtracted) RV time series and the $S_{HK}$ index, with $S_{HK} \gtrsim 0.2$, consistent with significant activity for stars of similar color to Kepler-1656 (Isaacson & Fischer 2010, Figure 9). We show the residual RV data as a function of $S_{HK}$ in the right panel of Figure 2. We computed a Pearson $r$ coefficient of 0.51 for this data set, suggesting a modest correlation; however, this is not uncommon for stars of this type and does not necessarily imply an activity-induced RV trend (Wright et al. 2008). Moreover, the data have a median $S_{HK}$ of 0.15, similar to low-activity stars of similar color. $S_{HK}$ values within this range are consistent with jitter $\sigma_j \approx 2$ m s$^{-1}$, which cannot account for the full $\approx 7$ m s$^{-1}$ variability we see in our RVs. Thus, our observed RV trend is most likely due to an undetected companion—either a star, brown dwarf, or second planet—in the system.

### 2.3. Periodogram

We conducted a periodogram search for an outer planet in our RV data using RVSearch, an algorithm that iteratively searches for periodic signals in RV time series data (Rosenthal et al. 2021). We finely sampled a grid of outer planet periods ($P_c$) with $31 < P_c < 10,000$ days. For each $P_c$, RVSearch compares the maximum likelihood two-planet model to the single-planet model by evaluating the change in the BIC (see Schwarz 1978). Peaks in the periodogram thus correspond to the most plausible periods for the outer planet. A periodogram search for two-planet signals is shown in the top panel of Figure 3. We also plot the $\Delta$BIC threshold for false detections, corresponding to an empirical false alarm probability (eFAP) of 0.1%.

As can be seen in the top panel of Figure 3, the BIC strongly favors the long-period candidate peak over the single-planet model with $\Delta$BIC = 79. This model is also favored over other potential two-planet models (i.e., other peaks in the periodogram) with $\Delta$BIC $\gtrsim 30$, and lies above the false detection threshold. We find no additional periodic signals after subtracting the two-planet signal from our RVs (i.e., considering the three- versus two-planet case), as shown in the bottom panel of Figure 3. Thus, we conclude that the long-term trend in our RVs is due to a second planet in the system, Kepler-1656c, whose orbital parameters correspond to the $P_c \sim 1920$ day peak in the periodogram.

Figure 4 shows the best-fit two-planet Keplerian model of the system, including both Kepler-1656b and Kepler-1656c.[9] We summarize our adopted system parameters from this model in

---
[9] We note that there is an outlier in our RV data that does not fit the model. We examined the pipeline reduction and spectrum associated with this RV point, but do not find any anomalies. Thus, we include the outlier in our analysis.

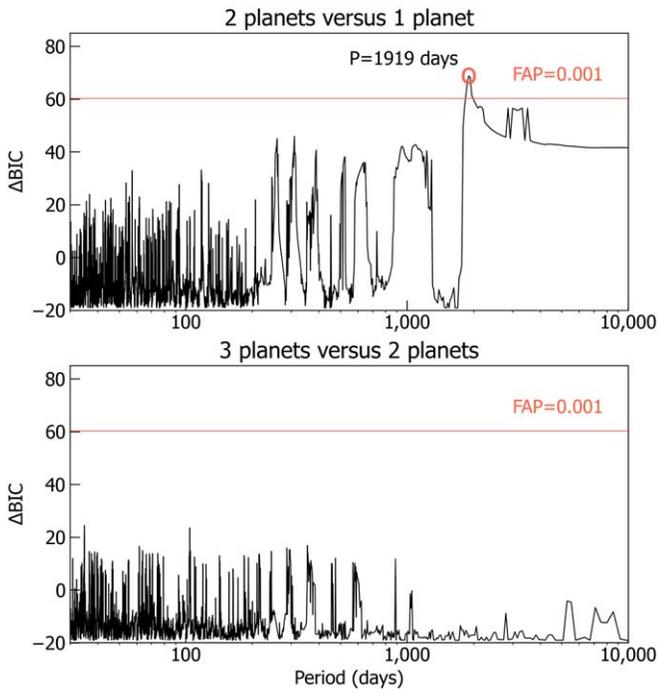

**Figure 3.** Top: change in Bayesian Information Criterion (BIC) comparing two-planet models and single-planet models. The trial period of the two-planet model is shown on the horizontal axis. The BIC favors $P_c \sim 1920$ days, corresponding to a second planet in the system. The BIC associated with eFAP = 0.001, above which peaks are considered true planets, is overplotted in red. Bottom: the same as the top panel, but comparing three-planet models to two-planet models. No peaks lie above the eFAP = 0.001 threshold, indicating no additional periodic signals are detected.

Table 2. We find that Kepler-1656c is a giant planet with $M_c = 0.43 \pm 0.10 \, M_{\rm jup}$, semimajor axis $a_c = 3.057^{+0.059}_{-0.052}$ au, and eccentricity $e_c = 0.53^{+0.064}_{-0.08}$. For the remainder of our analysis, we explore a parameter space in the vicinity of Kepler-1656c (see Figure 1) to uncover the origins of Kepler-1656b's high eccentricity and to understand the dynamical evolution of sub-Saturns with outer companions more generally.

## 3. Dynamical Model

Our observations in Section 2 revealed an outer planetary companion to Kepler-1656b. Gravitational perturbations from this companion should have a clear signature on the evolution and orbit of the inner planet. Thus, we can simulate the dynamical evolution of the Kepler-1656 system from an array of initial configurations to investigate formation scenarios that are consistent with the present-day properties.

An outer companion can excite a planet's eccentricity via dynamics that arise from a hierarchical three-body configuration—that is, a tight inner orbit of two bodies and a long-period outer orbit of a third body around the inner orbit's center of mass (see Figure 5 for visual depiction). In this case, the inner and outer orbits torque each other and exchange angular momentum over timescales much longer than the planets' orbital periods. This exchange produces long-term eccentricity and inclination oscillations known as the eccentric Kozai–Lidov (EKL) mechanism (Kozai 1962; Lidov 1962; Naoz et al. 2016). In this regime, the secular (i.e., long-term, orbit-averaged) evolution of the system is chaotic and the planet can undergo extreme eccentricity oscillations (e.g., Li et al. 2014a). This evolution can lead to interesting dynamics, such as engulfment of the inner planet, retrograde orbits, or high-eccentricity migration (e.g., Naoz et al. 2011, 2012; Lai et al. 2018; Vick et al. 2019). For example, hot Jupiters on orbits that are misaligned with their host star's spin can be attributed to this mechanism (Naoz et al. 2011, 2012; Albrecht et al. 2012; Anderson et al. 2016; Storch et al. 2017; Stephan et al. 2018).

In this section, we outline the physical setup and relevant dynamical processes at play for the Kepler-1656 system—that is, a two-planet system with a hierarchical orbital configuration subject to EKL behavior and other dynamical effects which we outline below. We then describe the underlying physics and initial conditions of our dynamical simulations.

### 3.1. Physical Setup and Relevant Timescales

Kepler-1656b's orbit together with the distant companion forms a hierarchical triple system in which Kepler-1656b orbits on a tight inner orbit with respect to the companion. In this section, we examine the general dynamical landscape of hierarchical two-planet systems like Kepler-1656 by comparing timescales associated with various physical processes that are at play. In particular, we consider hierarchical triple systems consisting of inner planets with outer companions in the vicinity of parameter space where Kepler-1656c resides (i.e., the outlined region of Figure 1) to capture a dynamical picture of Kepler-1656b and similar sub-Saturns with distant companions.

For an inner companion of mass $M_b$ and outer planet of mass $M_c$ orbiting a host star of mass $M_\star$ the EKL-induced eccentricity oscillations occur on a characteristic timescale, hereafter $t_{\rm EKL}$ (e.g., Antognini 2015):

$$t_{\rm EKL} = \frac{16}{30\pi} \frac{M_\star + M_b + M_c}{M_c} \frac{P_c^2}{P_b(1-e_c^2)^{3/2}}$$

$$\sim 8.3 \times 10^4 {\rm yr} \left(\frac{M_\star}{M_\odot}\right)\left(\frac{M_c}{M_{\rm jup}}\right)^{-1}$$

$$\times \left(\frac{P_b}{30 \, {\rm days}}\right)^{-1}\left(\frac{P_c}{5 {\rm yr}}\right)^2 \left[1 - \left(\frac{e_c}{0.5}\right)^2\right]^{-3/2}, \quad (1)$$

where $P_b$ and $P_c$ are the inner and outer orbital periods and $e_c$ is the eccentricity of the outer planet. Here we consider the lowest order level of secular approximation, namely the quadrupole level, which describes the shortest timescale over which the eccentricity oscillates (e.g., Naoz et al. 2013a). We plot $t_{\rm EKL}$ versus inner planet semimajor axis ($a_b$) for outer companions consistent with our observations as a blue shaded band in Figure 6. In our simulations, we also account for the fact that these oscillations may be modulated by the octupole-level order of the secular approximation, which can lead to more chaotic evolution and larger amplitudes (Naoz et al. 2016).

In addition to the EKL mechanism, there are several potentially relevant effects that may contribute to the evolution of the systems we consider. In particular, general relativity (GR) precession, as well as tidal effects like circularization and shrinking, all contribute to the secular

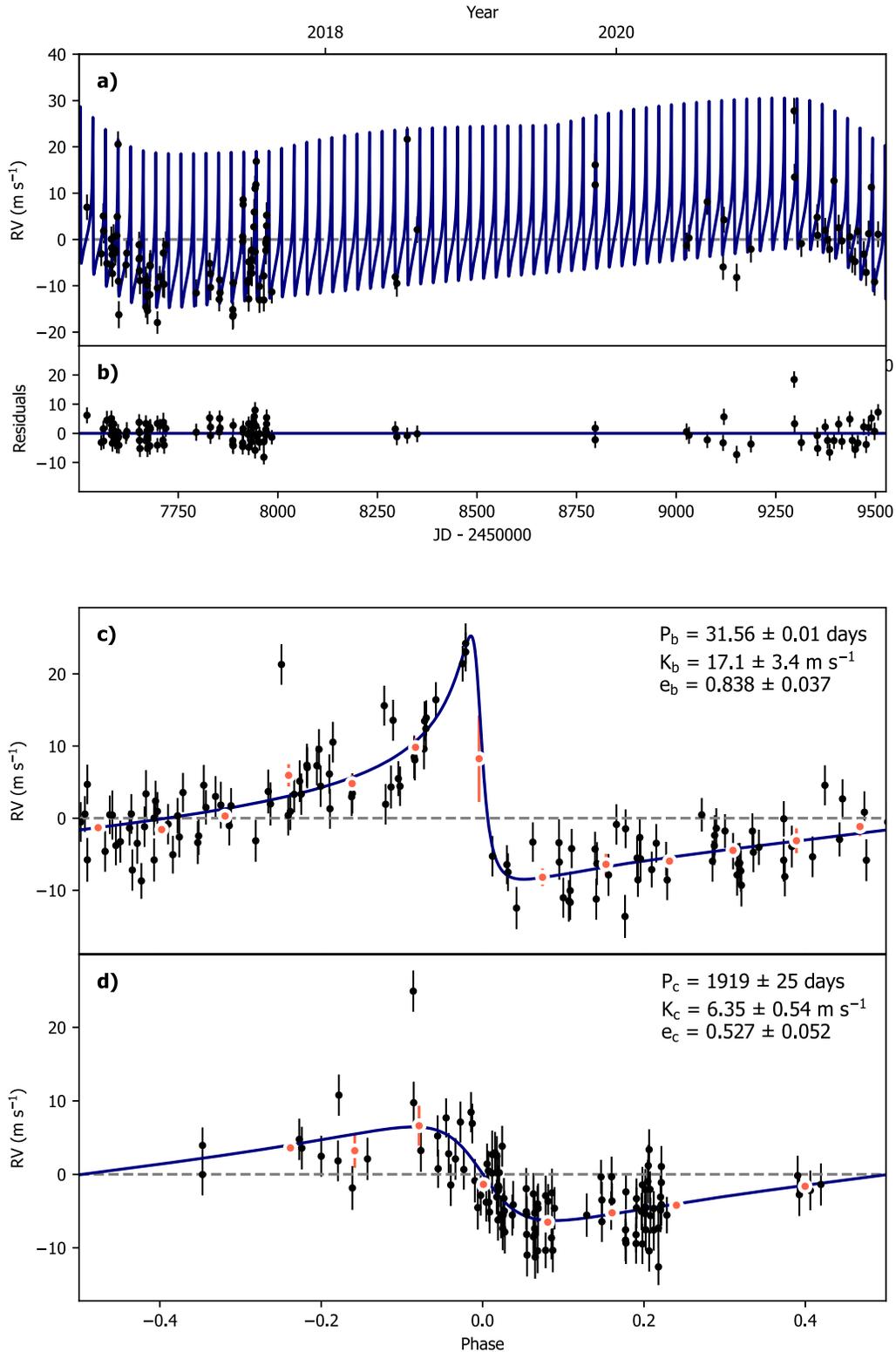

**Figure 4.** Adopted a two-planet Keplerian model for the Kepler-1656 system. (a) Black points are RVs and the blue line shows our maximum a posteriori model. (b) Model residuals are shown in black. (c) Observations are phase-folded over the inner planet period and the phase-folded Keplerian model is overplotted. RVs binned in 0.08 units of orbital phase are shown in red. (d) Observations are phase-folded over the outer planet period, phase-folded Keplerian model is overplotted.

dynamical evolution of the Kepler-1656 system. While the EKL mechanism excites eccentricity, the aforementioned processes tend to suppress it (e.g., Ford et al. 2000; Fabrycky & Tremaine 2007; Naoz et al. 2013). For instance, the GR precession of the inner planet orbit occurs on a typical timescale of

**Table 2**
Kepler-1656 System Parameters

| Parameter | Value | Notes |
|---|---|---|
| Stellar | | |
| $M_\star$ ($M_\odot$) | $1.03 \pm 0.04$ | A |
| $R_\star$ ($R_\odot$) | $1.10 \pm 0.13$ | A |
| age (Gyr) | $6.31^{+2.1}_{-2.9}$ | A |
| $v \sin i$ (km s$^{-1}$) | $2.8 \pm 1.0$ | B |
| Planet b | | |
| $P_b$ (days) | $31.562 \pm 0.011$ | C |
| $T_{c,b}$ (days) | $2455011.47 \pm 0.90$ | C |
| $e_b$ | $0.838^{+0.045}_{-0.029}$ | C |
| $\omega_b$ | $0.922^{+0.073}_{-0.095}$ | C |
| $K_b$ (m s$^{-1}$) | $17.1^{+5.3}_{-1.6}$ | C |
| $a_b$ (au) | $0.1974 \pm 0.0026$ | D |
| $M_b \sin i$ ($M_\oplus$) | $47.8^{+6.2}_{-3.3}$ | D |
| $M_b \sin i$ ($M_{\rm jup}$) | $0.150^{+0.019}_{-0.010}$ | D |
| $M_b$ ($M_\oplus$) | $47.8^{+6.2}_{-3.3}$ | D,E |
| $M_b$ ($M_{\rm jup}$) | $0.15 \pm 0.02$ | D,E |
| Planet c | | |
| $P_c$ (days) | $1919^{+27}_{-24}$ | C |
| $T_{c,c}$ (days) | $2459461^{+24}_{-26}$ | C |
| $e_c$ | $0.527^{+0.050}_{-0.054}$ | C |
| $\omega_c$ | $1.53^{+0.24}_{-0.28}$ | C |
| $K_c$ (m s$^{-1}$) | $6.35^{+0.56}_{-0.52}$ | C |
| $a_c$ (au) | $3.053 \pm 0.049$ | D |
| $M_c \sin i$ ($M_\oplus$) | $107.2 \pm 10.2$ | D |
| $M_c \sin i$ ($M_{\rm jup}$) | $0.337 \pm 0.032$ | D |
| $M_c$ ($M_\oplus$) | $126.4 \pm 28.9$ | D,F |
| $M_c$ ($M_{\rm jup}$) | $0.40 \pm 0.09$ | D,F |

**Note.** A: Johnson et al. (2017); B: Petigura et al. (2017b); C: This work, adopted the Keplerian model; D: This work, derived; E: Gaussian priors imposed on $M_b \sin i$ based on our RV analysis and $i = 89.31 \pm 0.51$ reported in Brady et al. (2018); F: Gaussian prior imposed on $M_c \sin i$ based on our RV analysis, and uniform prior imposed on $\cos(i)$.

$$t_{\rm GR} = 2\pi \frac{a_b^{5/2} c^2 (1-e_b^2)}{3 G^{3/2} (M_\star + M_b)^{3/2}}$$

$$\sim 2.2 \times 10^5 \text{yr} \left(\frac{M_\star}{M_\odot}\right)^{-3/2} \left(\frac{a_b}{0.2 \text{ au}}\right)^{5/2}$$

$$\times \left[1 - \left(\frac{e_b}{0.8}\right)^2\right], \quad (2)$$

where $G$ is the gravitational constant and $c$ is the speed of light. This precession suppresses EKL eccentricity excitations if $t_{\rm GR}$ is shorter than $t_{\rm EKL}$ (i.e., the upper left region of Figure 6).[10]

Given Kepler-1656b's high eccentricity it is also important to compare $t_{\rm EKL}$ to the shrinking and circularization timescales ($t_{\rm shrink}$, and $t_{\rm circ}$, respectively) of the inner orbit due to tides. We estimate these timescales for inner planet orbit due to equilibrium tides (e.g., Eggleton et al. 1998; Eggleton & Kiseleva-Eggleton 2001), which for $e_b \sim 0.8$ can be approximated as:

$$t_{\rm shrink} \sim \frac{t_{v,b} a_b^8 M_b^2 (1-e_b^2)^{15/2} f_T(e_b)}{162(1+2k_{s,b})^2 R_b^8 M_\star (M_\star + M_b) e_b^2}$$

$$\sim 4.1 \times 10^{10} \text{yr} \left(\frac{M_\star}{M_\odot}\right)^{-2} \left(\frac{a_b}{0.2 \text{ au}}\right)^8 \left(\frac{M_b}{50 M_\oplus}\right)^2$$

$$\times \left(\frac{R_b}{5 R_\oplus}\right)^{-8} \left[1 - \left(\frac{e_b}{0.8}\right)^2\right]^{15/2} \left(\frac{e_b}{0.8}\right)^{-2} \times f(e_b) \quad (3)$$

and

$$t_{\rm circ} \sim \frac{t_{v,b} a_b^8 M_b^2 (1-e_b^2)^{13/2} f_T(e_b)}{81(1+2k_{s,b})^2 R_b^8 M_\star (M_\star + M_b)}$$

$$\sim 1.5 \times 10^{11} \text{yr} \left(\frac{M_\star}{M_\odot}\right)^{-2} \left(\frac{a_b}{0.2 \text{ au}}\right)^8 \left(\frac{M_b}{50 M_\oplus}\right)^2$$

$$\times \left(\frac{R_b}{5 R_\oplus}\right)^{-8} \left[1 - \left(\frac{e_b}{0.8}\right)^2\right]^{13/2} \times f(e_b), \quad (4)$$

where $R_b$ is the radius of the inner planet, $t_{v,b}$ is the viscous timescale of the inner planet, $k_{s,b}$ is the classical apsidal motion constant, and we define $f(e) \equiv 1/(1 + 15/4 e^2 + 15/8 e^4 + 5/64 e^6)$. Note that these are approximate scalings for the Kepler-1656 system in its current configuration—for our simulations, we use the full equations derived from Eggleton & Kiseleva-Eggleton (2001) which also depend on $\Omega_{s,b}$, the spin rate of the inner planet (see Naoz et al. 2016, Equation (86)–(90)). We adopt nominal values of $k_{s,b} = 0.25$ and $t_{v,b} = 1.5$ yr in these equations. Our nominal $t_{v,b}$ corresponds to a tidal quality factor of $Q_b \approx 1.8 \times 10^6$ for the planet, where $t_{v,b}$ is related to $Q_b$ by following the equilibrium tide formalism from Naoz et al. (2016):

$$Q_b = \frac{4}{3} \frac{k_{s,b}}{(1+2k_{s,b})^2} \frac{GM_b}{R_b^3} \frac{P_b t_{v,b}}{2\pi}. \quad (5)$$

We show $t_{\rm shrink}$ and $t_{\rm circ}$ timescales for a range of $a_b$ in Figure 6, where we consider both tides due to the planet and due to the star (see label). For the purpose of this figure we spin periods between 1–30 days for both the planet and star (i.e., $\Omega_{s,b} = 0.03$–1 days$^{-1}$, shown as shaded bands). As can be seen, both the shrinking and circularization timescales are longer than the system's age for the planet's observed semimajor axis.[11]

We see from Figure 6 that Kepler-1656b's current orbit (vertical dashed line) lies in the regime where EKL and GR precession timescales are shortest, indicating these timescales will compete to dominate the overall system behavior. Additionally, for some companions (i.e., in the left-hand region of the blue band in Figure 6), $t_{\rm shrink}$ and $t_{\rm circ}$ are similar in magnitude to $t_{\rm EKL}$. Accordingly, for initial configurations where Kepler-1656b orbits closer to the host star, tides may play a role in suppressing eccentricity excitations (note that these timescales are plotted for Kepler-1656b's current orbit with $e_b \approx 0.8$; however, inner planets with higher eccentricities may have shorter tidal timescales). All this is to say that different companions will either induce or suppress eccentricity oscillations for Kepler-1656b.

---
[10] Although, in some cases, when the GR precession timescale is of the order of the quadrupole timescale, it can destabilize the resonance and potentially retrigger high-eccentricity excitations (e.g., Naoz et al. 2013; Hansen & Naoz 2020).

[11] We note that equilibrium tides tend to underestimate the efficiency of the tides compared to chaotic dynamical tides for sufficiently large eccentricity (Vick et al. 2019). Thus, for Kepler-1656b's eccentricity, the chaotic dynamical tides may result in more close-in, tidally locked systems than predicted by equilibrium tides, as we discuss in more detail in Section 4.1

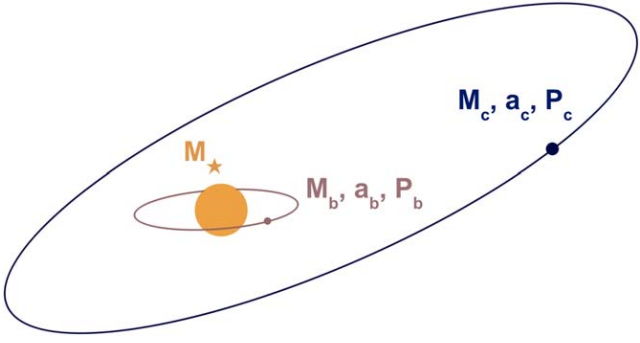

**Figure 5.** Schematic of a hierarchical triple system containing Kepler-1656b ($M_b$) on a tight inner orbit and a companion ($M_c$) on a distant, possibly inclined outer orbit around their host star ($M_\star$). Note that the sizes of the planets and their orbits are not drawn to scale.

We model the secular evolution of the system (up to the octupole-level of approximation) taking all of the aforementioned physical processes into account, i.e., EKL, GR, and tidal circularization and shrinking. We also include stellar evolution, which traces the host star's spin. Stellar evolution can play an important role in the dynamics of the planet. For example, an expanding star can engulf or circularize the inner planet, while a star undergoing mass loss can widen its orbit (e.g., Dobbs-Dixon et al. 2004; Stephan et al. 2017, 2018, 2020b, 2020a). Our simulations compute the secular evolution of the inner and outer orbits over time from a suite of initial configurations which we describe in Section 3.2.

### 3.2. Initial Conditions

In all of our simulations, we fix the host star and inner planet parameters to values reported in Table 2 and sample from the parameter space surrounding our best-fit values for planet c. We set the mass and radius for the host star and inner planet to $M_b = 48.6 \, M_\oplus$, $R_b = 5.02 \, R_\oplus$, $M_\star = 1.03 \, M_\odot$, and $R_\star = 1.10 \, R_\odot$. We considered candidate outer planets in the vicinity of Kepler-1656c with $P_c = 600 - 4000$ days, corresponding to a uniform distribution of $a_c = 1.4 - 5$ au, and $M_c$ from a log-uniform distribution within the planetary mass regime of $0.1 \, M_b < M_c < 2 \, M_{jup}$. We sample the mutual inclination from an isotropic distribution (i.e., uniform distribution for $\cos i_{bc}$) and $e_c$ from a uniform distribution of (0, 1). These initial conditions are depicted in Table 3. To explore the origins of Kepler-1656b's eccentricity, we place Kepler-1656b at a near-circular orbit initially ($e = 0.01$).

We fix the viscous timescale of the star and planet to $t_{v,\star} = t_{v,b} = 1.5$ yr in our simulations. We note that while more detailed modeling of tidal effects is valuable for this system, it is beyond the scope of this paper. We model the secular evolution of the Kepler-1656 system by numerically solving the octupole-level Hamiltonian for the hierarchical triple following Naoz et al. (2013a), folding in GR effects for the inner and outer orbit (Naoz et al. 2013), and tidal effects (following Eggleton & Kiseleva-Eggleton 2001). We account for the star's stellar evolution using the SSE evolution code (Hurley et al. 2000), which is important, particularly in accounting for the stellar spin's evolution (this combined code was tested and applied to various ranges of astrophysical systems; see for example Naoz 2016a; Stephan et al. 2016, 2018). To account for the mentioned uncertainty in

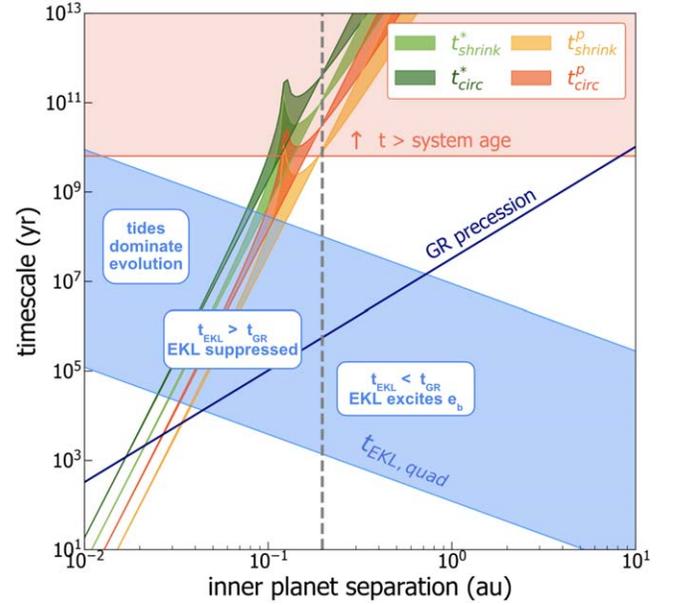

**Figure 6.** Relevant dynamical timescales for Kepler-1656b in the presence of companions similar to Kepler-1656c are plotted as a function of the inner planet orbital separation, $a_b$. The shaded blue band represents the range of eccentricity oscillation timescale $t_{EKL}$ for outer companions in the parameter space surrounding Kepler-1656c, with $0.1 \, M_b < M_c < 2 \, M_{jup}$, $P_2 = 600-4000$ days ($a_c = 1.4 - 5$ au), and $e_c = 0 - 0.9$. For the tidal shrinking and circularization timescales, we consider a range of planet and star spin periods from 1-30 days. The upper boundary of the blue region is set by $a_c = 5$, $M_c = 0.1 \, M_b$, $e_c = 0$, and the lower boundary is set by $a_c = 1.4$, $M_c = 2 \, M_{jup}$, $e_c = 0.9$.

**Table 3**
Simulation Overview

| Parameter | Set 1 | Set 2 | Distribution |
|---|---|---|---|
| $N_{sim}$ | 800 | 1000 | … |
| $a_b$ | 0.1–10 au | 0.197 au | log-uniform (1), fixed (2) |
| $a_c$ | 1.4–5 au | "[a] | uniform |
| $M_b$ | 48.6 $M_\oplus$ | " | fixed |
| $M_c$ | 0.1 $M_b$-2 $M_{jup}$ | " | log-uniform |
| $e_b$ | 0.01[b] | " | fixed |
| $e_c$ | 0–1 | " | uniform |
| $i_{bc}$ | 0°–180° | " | uniform in $\cos(i_{bc})$ |

**Notes.**
[a] denotes the same values as in Set 1.
[b] not fixed at $e_b = 0$ exactly to avoid errors in our numerical calculations.

the tidal timescale as well as the large uncertainty of the system's age, we focus on timestamps within ±20% of the system age, $t_{sys} = 6.31$ Gyr. This window is conservative compared to the system error bars ($\Delta t \approx 3$ Gyr; see Johnson et al. 2017).

We ran two sets of simulations under the aforementioned initial conditions: one where we sample $a_b = 0.1 - 10$ au to allow for the possibility of migration, and one in which begins in situ at $a_b = 0.197$ au (see Table 3). The first set of 800 simulations (hereafter Set 1) samples $a_b$ from a log-uniform distribution between 0.1 au and a maximum value set by the following stability criterion:

$$a_b \leqslant \epsilon_{max} a_c \frac{1 - e_c^2}{e_c}, \quad (6)$$

where we fix $\epsilon_{max} = 0.1$ to be the maximum relative amplitudes of the octupole and quadrupole terms in the Hamiltonian. This criterion verifies that the system is hierarchical and that the secular approximation is valid (Naoz et al. 2016).[12]

Following the results of Set 1, we find that only the systems that began in situ (i.e., with $a_b = 0.197$ au initially) ended up in the regime that matches Kepler-1656b's observational constraints (see Section 4.1 for more details). This motivated us to run a second set of simulations that only considers systems beginning in situ with $a_b = 0.197$ au, hereafter Set 2. For Set 2, we draw from the same distributions as Set 1 for all parameters except $a_b$ and apply the aforementioned stability criterion by rejecting initial conditions for which $\epsilon > 0.1$ (see Equation (6)). We run 1000 of these systems, as summarized in Table 3.

## 4. Dynamical Analysis

### 4.1. Kepler-1656b Has Not Migrated

Results from Set 1 of our simulations (i.e., models in which Kepler-1656b was sampled at a range of initial semimajor axes) are plotted as a function of the planet's eccentricity and semimajor axis in the top panel of Figure 7. We overplot the region of eccentricity versus semimajor axes (hereafter "e-a") space that is consistent with our observations as a black rectangle. Here we define consistent as having a $a_b = 0.16 - 0.24$ au, $e_b = 0.67 - 1$ (note however that in some cases in the rectangle where $e_b \approx 1$, the planet plunges into the host star and is thus inconsistent with our observations, as discussed further in Section 4.2). For these simulations, only 2.0% ended up in the observed configuration at some point within 20% (1.2 Gyr) of the system's age. As seen in Figure 7, all of these models were in situ initially, providing early evidence that Kepler-1656b began its dynamical lifetime in situ.

The upper envelope of e-a space overplotted in Figure 7 shows the constant angular momentum track that planets undergoing high-eccentricity migration follow. For these planets, scattering, EKL, or secular chaos (Rasio & Ford 1996; Chatterjee et al. 2008; Naoz et al. 2011; Lithwick & Wu 2014, see) can cause tidal effects to dominate over perturbations from the outer companions. These tidal effects cause the inner planet to circularize and migrate inward. The initial angular momentum for these eccentric systems, with initial eccentricity $e_{b,i}$ and semimajor axis $a_{b,i}$, can be written as:

$$J^2 \propto a_{b,i}(1 - e_{b,i}^2) \approx 2a_{b,i}(1 - e_{b,i}), \quad (7)$$

since $e_{b,i} \to 1$ for an initially eccentric planet. Similarly, the final angular momentum can be written as:

$$J^2 \propto a_{b,f}(1 - e_{b,f}^2) \approx a_{b,f}, \quad (8)$$

since $e_{b,i} \to 0$ as the planet circularizes. The planet's orbit can continue to shrink without being tidally disrupted as long as its closest approach $r_{peri} = a_b(1 - e_b)$ is larger than its Roche limit, so the lowest possible periastron the planet can shrink to is $a_{b,f}(1 - e_{b,f}) \approx a_{b,f} \approx R_{Roche}$. Angular momentum is conserved in this process; thus, we can combine Equations (7) and (8) to get:

$$a_{b,i}(1 - e_{b,i}^2) \approx a_{b,f} = R_{Roche}. \quad (9)$$

We show this envelope as a solid black line in Figure 7 for $a_{final} = 0.01$ au. The shaded region around this line shows the same upper envelope for a range of $a_{b,f}$ within a factor of 2. As we can see, Kepler-1656b does not migrate from its initial semimajor axis in any of our simulations prior to reaching this upper envelope.

We also consider a second e-a upper envelope for Kepler-1656b set by chaotic tides between the planet's orbital energy and fundamental modes (i.e., "f-modes", e.g., Wu 2018; Vick et al. 2019). For close, eccentric orbits, a planet's orbital energy may be converted into internal fluid energy (a.k.a. "f-mode oscillations") via tidal stretching and compression at periastron passage, thereby causing the orbit to circularize over time. Wu (2018) describes the quantitative criterion for this chaotic tidal evolution to occur:

$$a_b(1 - e_b) \leqslant 0.02 \text{ au} \left(\frac{R_b}{1.1 R_{jup}}\right)^{2/21} \left(\frac{a_{b,i}}{1 \text{ au}}\right)^{5/42}$$
$$\times \left(\frac{Q'_{nl}}{0.5}\right)^{2/21} \left(\frac{P}{1.04 \times 10^4 s}\right)^{11/21}, \quad (10)$$

where $a_{b,i}$ is Kepler-1656b's initial separation, (i.e., 0.197 for the in situ case), $Q'_{nl}$ is a dimensionless form of the tidal integral (set to 0.5 for f-modes), and $P$ is the period of the prograde mode ($P = 3.4 \times 10^4$ s for a spin period 10 days).

We overplot this criterion as a dashed upper envelope to the e-a in Figure 7. We see here that our simulations do not reside in the regime where chaotic tides play a large role in the evolution of Kepler-1656b (i.e., the e-a region between the dashed and solid lines). Furthermore, we see from this figure that our simulations do not migrate prior to reaching the upper envelope set constant angular momentum. Thus, only systems that began in situ can reach the high eccentricity we observe. Motivated by these results, Set 2 requires Kepler-1656b to be in situ at the start of our simulations.

Set 2 produces results that are significantly more consistent with our observations, with 3.3% ending up in the observed configuration within 20% of the system's age (compared to 2.0% in Set 1). The absence of pre-envelope migration in both sets of simulations and the increased likelihood of matching our observations in Set 2 both suggest that the planet began its dynamical lifetime in situ. We stress, however, that this does not require the planet to have originally formed in situ—there are a number of different paths for Kepler-1656b to reach this configuration prior to its in situ EKL onset, which we discuss in Section 5.

### 4.2. Fate of Sub-Saturns with Distant Giant Companions

Figure 7 shows the evolution of all models in Set 1 (top row) and Set 2 (bottom row) of our simulations in e-a space, color-coded according to the final outcome for that system. In both rows, the left panel shows only the initial and final timestamps of each simulation, and the right panel shows all timestamps within 20% of the system's age. We overplot the constant angular momentum envelope in each panel, as well as the aforementioned diffusive tides criterion from Wu (2018; see Equations (9) and (10), respectively).

---

[12] To verify our stability condition, we calculated the timescale for the Kepler-1656 system to become unstable independently using the stability metric from E. Zhang et al. (2022, in preparation). We find that systems with $\epsilon > 0.1$ become unstable in timescales much smaller than the EKL timescale (Equation (A1)) by this metric, confirming that our stability criterion is reasonable.

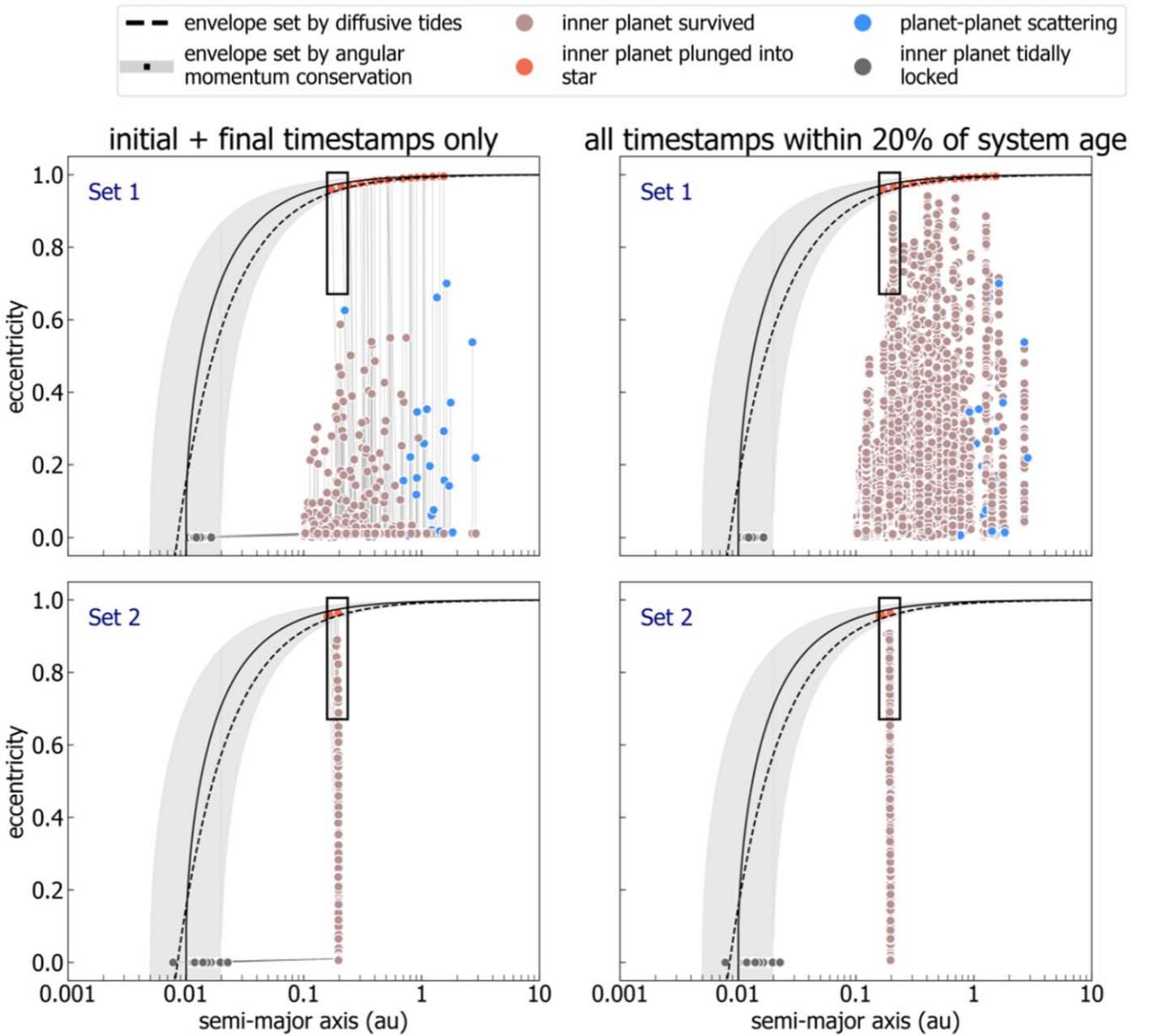

**Figure 7.** Eccentricity and semimajor axis of simulated planet b's with planet c parameters drawn according to the distributions described in Table 3. The top row shows selected timestamps from simulations Set 1, with just initial and final timestamps at $t = 0$ and $t = 7.5$ $Gyr$ connected by gray lines on the left, and all timestamps within 20% of the system's age (i.e., $t = 5.4$–$7.5$ Gyr) on the right. The bottom row shows the same for simulations Set 2. In all panels, the points are colored according to the simulation outcome. The upper envelope set by constant angular momentum (Equation (9)) is plotted in black for a range of $a_{b,f}$ within a factor of 2 of 0.01 au (shaded region), as well the upper envelope set by chaotic tides (see Wu 2018, Equation (20)). The region of the diagram that lies within 20% of the observed semimajor axis and eccentricity values reported in Brady et al. (2018) is outlined in black.

In general, there are four possible outcomes for our simulations:

1. *Inner planet crosses the star's Roche limit:* If the eccentricity of the inner planet is extremely high, as can be the case for the EKL mechanism (e.g., Li et al. 2014b), the orbit can fall within the star's Roche lobe at closest approach, causing the planet to plunge into its host star. We find that 0.8% (8.0%) of systems are plunged into their host in Set 2 (1) of our simulations. We mark these systems in red in Figure 7, and as is depicted, they are confined to the upper part of the angular momentum envelope.

   Crossing the host star's Roche limit is a typical outcome of the EKL mechanism (Naoz et al. 2012), and can have ramifications on the star's spin (e.g., Soker & Harpaz 2000; Metzger et al. 2012; Qureshi et al. 2018; Stephan et al. 2020b) and lithium abundance (Aguilera-Gómez et al. 2016; Bharat Kumar et al. 2018). We see evidence for this in pre-engulfment hot Jupiters and ultra-short-period planets (USPs) as well as post-engulfment white dwarfs (see Jura et al. 2009; Zuckerman et al. 2010; Maciejewski et al. 2016; Stephan et al. 2018).

   The orbits of the systems highlighted in Figure 7 are short-lived. Additionally, these systems tend to have closer and more eccentric companions ($e \approx 0.5$) and slightly misaligned mutual inclinations ($i_{bc} \sim 40^{o}$), which are generally less stable than their retrograde counterparts

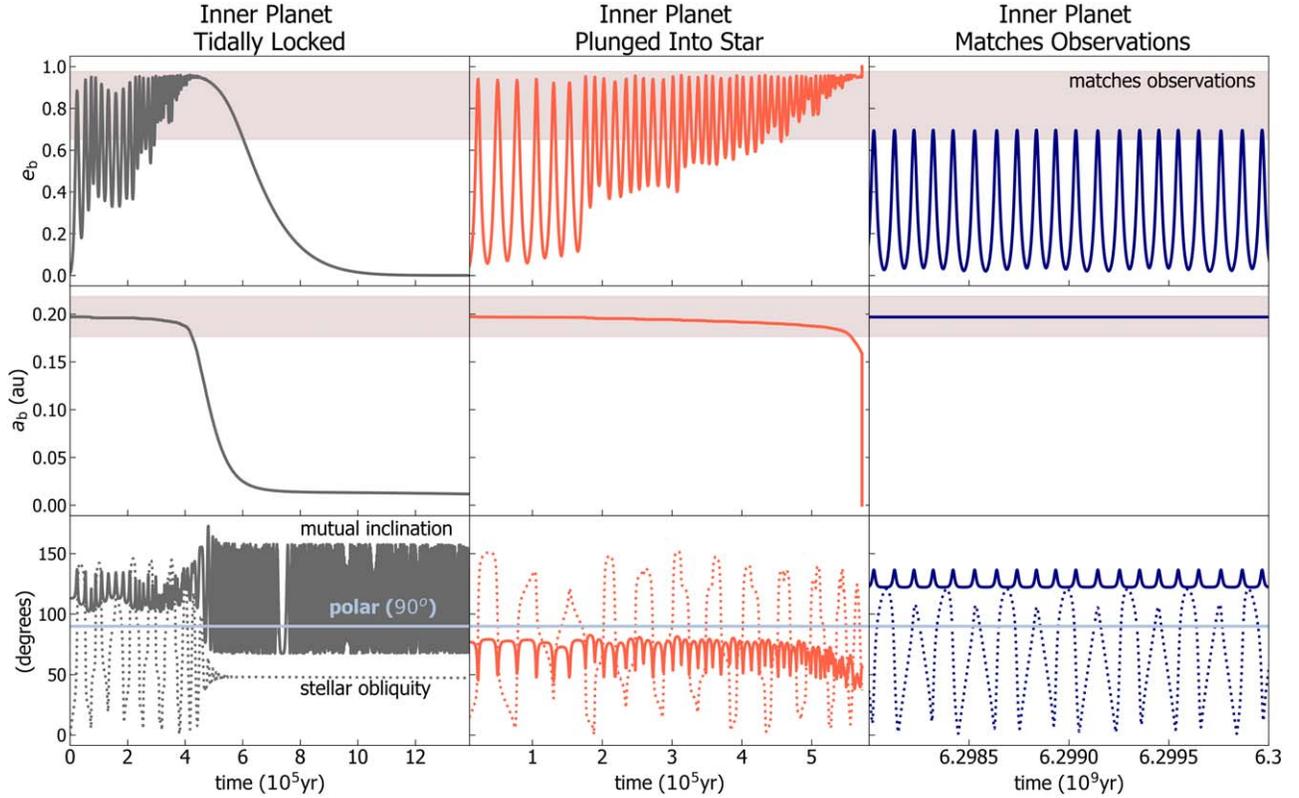

**Figure 8.** Evolution of three simulations with qualitatively different outcomes. Left: for a system with $M_c = 1.52\,M_{jup}$, $a_c = 2.52$ au, $e_c = 0.64$, $i_{bc} = 112.96°$ initially, the inner planet becomes tidally locked after $1.2 \times 10^6$ yr. Middle: for a system with $M_c = 1.28\,M_{jup}$, $a_c = 1.88$ au, $e_c = 0.51$, $i_{bc} = 76.93°$ initially, the inner planet crosses the host star's Roche limit after $5 \times 10^6$ yr. Right: for $M_c = 1.09\,M_{jup}$, $a_c = 2.09$ au, $e_c = 0.014$, $i_{bc} = 121.65°$ initially, we reproduce our observations for a sustained, eccentric orbit for the inner planet past $6.3 \times 10^9$ yr.

(e.g., Innanen et al. 1997). This is evident in Figure 9 which shows the distribution of companion parameters for the subset of simulations that plunge into their host star, as well as subsets for the other three simulations outcomes (as also highlighted in the density maps in Appendix B). They are also consistent with more massive outer companions (Figure 9, middle panel of row 2) which will tend to simultaneously perturb the inner planet and resist torque induced by it. An example of the evolution of such a system is shown in Figure 8. These outcomes suggest that sub-Saturns engulfed by their star's Roche lobe are more likely to arise from slightly misaligned ($i_{bc} \sim 40°$) two-planet systems with eccentric, massive ($M \geqslant 1\,M_{jup}$) outer companions.

2. *Inner planet is tidally locked:* When the eccentricity is excited to high values but the planet remains outside the star's Roche limit, tidal effects start to dominate and the planet's semimajor axis shrinks and circularizes, settling in a tidally locked configuration for the inner orbit. For our simulations, we found that 10.3% (15.3%) of systems in Set 2 (1) were tidally locked within the 7.5 Gyr of our simulations. They are marked in gray in Figure 7 and are connected to their initial $e$-$a$ configurations in the left panel. We also show the evolution of one of these tidally locked systems in the left panel of Figure 8.

The process of tidal locking has been discussed extensively in the literature (e.g., Fabrycky & Tremaine 2007; Wu et al. 2007; Naoz et al. 2012; Petrovich 2015). Along with stellar binary companions, planetary companions can also torque the inner planet and induce tidal migration and locking (Naoz et al. 2011; Teyssandier et al. 2013; Petrovich & Tremaine 2016). It has been shown that mutual inclination between nominal Kozai angles of $\sim 40°$ and $\sim 140°$ is required to reach the high eccentricity needed for tides to shrink and circularized the orbit. We see here (top row, left panel) that initially large inclinations, along with moderate outer orbit eccentricity, are indeed needed to produce tidally locked orbits (see also Appendix B). These results imply that warm, tidally locked sub-Saturns in two-planet systems are born out of configurations in which the outer companion is eccentric and mutually inclined with respect to the inner orbit. We classify outer companions that cause their inner planet to tidally lock as "strong companions" since they have drastic effects on the system's orbital architecture (see Section 5 for a detailed discussion).

3. *Planet–planet scattering:* For configurations in which the inner and outer orbits are tightly packed, the system can become dynamically unstable or only marginally hierarchical (e.g., Bhaskar et al. 2021). For our simulations, this instability can arise from one of two possible outcomes: (1) the orbital configuration becomes non-hierarchical (i.e., $a_{b,apo}/a_{c,peri} \leqslant (M_b/M_c)^{1/3}$), in which case the secular approximation breaks down and the inner planet is gravitationally captured by the outer planet (Naoz & Silk 2014), or (2) the planets orbit within a factor of 5 of their mutual Hill radius, thus becoming dynamically unstable over timescales of $\sim 10^9$ yr (Chatterjee et al. 2008). These scenarios cause the hierarchical approximation in our simulations to break down, thus we do not model the dynamical evolution beyond this point. Instead, we classify both of these events as planet–planet scattering outcomes, whereby the gravitational interactions between the two planets lead to dynamical instability and possible collisions or ejections (e.g., Rasio & Ford 1996; Chatterjee et al. 2008). For our simulations, we find that 0.1% (6.0%) of systems in Set 2 (1) produce this outcome. As shown in Figure 7, simulations in which Kepler-1656b starts further out (i.e., simulations in Set 1) are more likely to scatter because the inner planet is closer, on average, to the companion from the start.[13]

Planet–planet scattering has been suggested to explain both individual planetary systems and population-level trends. For example, the misaligned orbits of KOI-13.01, HD 147506b, and HAT-P-7b, as well as multiple eccentric orbits in the upsilon-Andromedae system, are thought to be the outcome of planet–planet scattering events (Ford et al. 2005; Winn et al. 2007; Narita et al. 2009; Barnes et al. 2011). Beyond this, demographic trends such as the observed distribution of planet eccentricities, correlations between host star metallicity and orbital period, and orbital properties of hot Jupiters can be explained at least in part by planet–planet scattering (Ford & Rasio 2008; Dawson & Johnson 2018; Petigura et al. 2018b). We also speculate that perhaps a planet–planet scattering event resulted in the in situ configuration we propose for the onset of EKL effects in the Kepler-1656 system (see Section 5 for details). This speculation is consistent with the recent study of the HR 5183b system (Mustill et al. 2021).

4. *Inner planet survives and matches our observations:* In some cases, the eccentricity of the inner planet is excited via the EKL mechanism to values that are just below the upper envelope of the *e-a* diagram. The result of this evolution is a significantly eccentric inner orbit ($e_b \geqslant 0.67$, within 20% of the observed value for Kepler-1656b),[14] but still low enough to be stable over ∼$10^9$ year timescales without being subject to the aforementioned tidal effects. Orbits like this are consistent with our observations of Kepler-1656b and are the outcome for 3.3% (2.0%) of our simulations in Set 2 (1) (see Figure 8 for an example).

Stable, eccentric orbits like Kepler-1656b's are a predicted outcome for systems subject to EKL-dominated evolution (e.g., Naoz et al. 2011, 2012; Li et al. 2014b; Petrovich 2015; Petrovich & Tremaine 2016). In these cases, torque exchanged by the inner and outer orbits causes the inner planet's eccentricity to oscillate with brief excursions to high values (see Figure 8 for detailed evolution). For simulations that remain below the upper *e-a* envelope, these oscillations are sustained over timescales up to and past $t = 6.3 \, Gyr$, thus reproducing our observations at the system's reported age.

It is also important to note that even for systems that match our observations, Kepler-1656b spends a large fraction of its lifetime in much less eccentric orbits due to the nature of its eccentricity oscillations, as can be illustrated in the upper right panel of Figure 8. As such, it is possible that several other known, less eccentric exoplanets are undergoing similar high-amplitude eccentricity oscillations induced by outer companions, but are observed during a period of lower eccentricity. We classify companions that induce these in situ eccentricity oscillations as "gentle companions" because they are relatively well-behaved compared to their more disruptive "strong companion" counterparts, as discussed further in Section 5. Kepler-1656c is an example of one such companion, and drives Kepler-1656b to oscillate between low and high values throughout its lifetime.

5. *Inner planet survives, but does not match our observations:* The majority of our simulations (68.7% and 85.5% for Set 1 and 2, respectively) do not fall in any of the categories described above, and instead simply retain less eccentric ($e_b < 0.67$) orbits for the duration of our simulations (see Figure 7). Still, these systems are valuable to our analysis because their initial conditions can be ruled out as possible origins scenarios to Kepler-1656b since they are unable to reproduce our observations. Additionally, their low-to-moderate eccentricities and large distance from the upper envelope place them in a distinct region of *e-a* space compared to their more eccentric counterparts described above. Thus, planets in this region of parameter space may be signatures of distant companions that are exciting their eccentricities to values lower than that of Kepler-1656b. We discuss this possibility in more detail in Section 5.

The fact that a subset of our EKL simulations is consistent with our observations of Kepler-1656b indicates that the EKL mechanism is a plausible origin scenario for Kepler-1656bs eccentric orbit. Moreover, the range of possible outcomes in our simulations suggests a diverse set of dynamical histories for sub-Saturns with distant giant companions. We discuss these histories and their potential signatures on the *e-a* distribution of exoplanets in Section 5.

### 4.3. Properties and Influence of the Outer Planet

Our simulations reveal that a subset of our sampled companions can reproduce the observed orbital configuration. Thus, we can search for commonalities in this subset to constrain both the evolution and observable properties of Kepler-1656c. Since Set 2 is more consistent with our observations, we focus on just this set for the remainder of our analysis of Kepler-1656 and return to our discussion of Set 1 in Section 5 to make generalizations about the sub-Saturn population.

The simulations that reproduce Kepler-1656b's orbit evolve from a unique set of orbital configurations (see Figure 9, right column), allowing us to uncover possible origin scenarios for the system. In particular, models that match our observations have identical initial and final distributions for the companion semimajor axis ($a_c$). This suggests that $a_c$ does not change over time, and much like its inner companion, Kepler-1656c likely began its dynamical lifetime in situ.

---

[13] We do not include these systems in Figure 9 because only one simulation from Set 2 has this outcome. This system had an outer companion with $M_c = 0.04 \, M_{jup}$, $a_c = 2$ au, $e_c = 0.42$, $i_{bc} = 105°$ initially.

[14] Note that this 20% is larger than the reported error bars in Table 2. This relaxed constraint is to account for both the finite time sampling in our simulations and the chaotic nature of the EKL mechanism (e.g., Naoz et al. 2016)

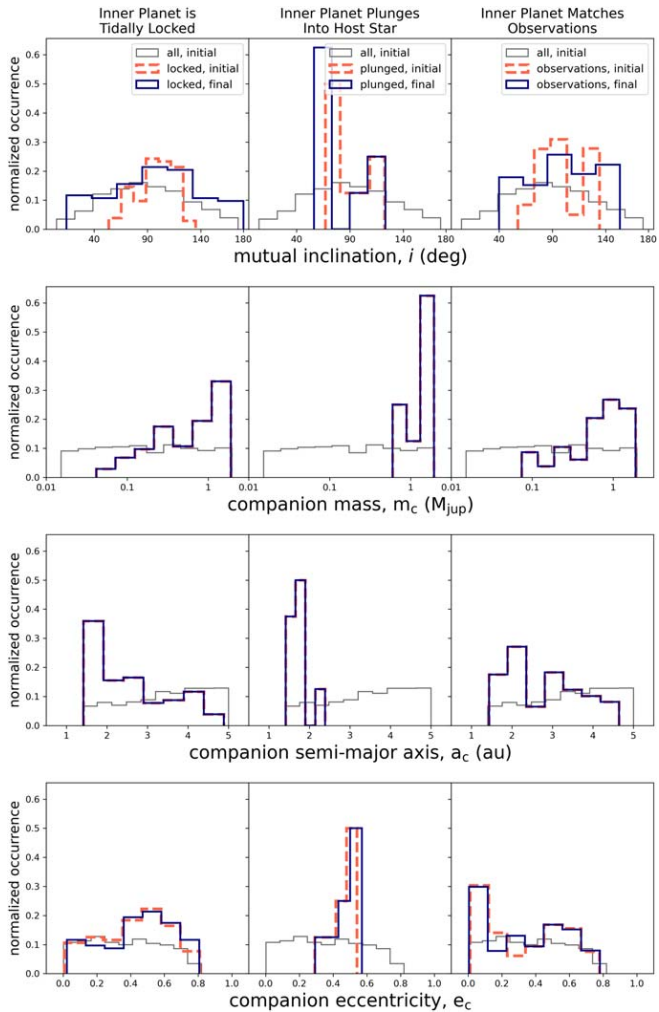

**Figure 9.** Normalized distributions are plotted for various orbital outer planet parameters in Set 2 of our simulations. The models are divided into the following subsets: models in which the inner planet becomes tidally locked (left), models in which the inner planet plunges into its host star (middle), and models in which the inner planet is consistent with our observations (right). For all panels, the initial distribution of our entire set of models is shown in black. Initial and final distributions of different subsets are overplotted in red and blue, respectively.

Figure 9 also shows that although Kepler-1656b and Kepler-1656c can end up with any mutual inclination within $i_{bc} = 40-150°$, they are born out of a slightly more narrow distribution of $i_{bc} = 60°-130°$. We also see that Kepler-1656c can be anywhere from 0–0.8 initially. These initial conditions are consistent with predictions conditions required for EKL onset—that is, an eccentric outer orbit and significant mutual inclination between the inner and outer orbits (Naoz et al. 2016). In fact, our simulations show that Kepler-1656b's dynamical behavior is driven by quadrupole-level EKL oscillations (see Appendix A, Equation (A1)). These oscillations perturb the inner orbit just enough to excite it to its high eccentricity, but not enough to induce high-eccentricity migration. We note that these are not test-particle quadrupole level as the inner orbit's angular momentum is not conserved Naoz et al. (2011, 2013a). Further discussion on this dynamical behavior can be found in Appendix A.

Beyond this, commonalities in the simulations that reproduce our observations inform us about the companion's current orbital configuration. This is evident in Figure 10, which shows 2D distributions of the companion properties in our simulations, both initially (in tan) and today (i.e., when $t$ is within 20% of $t_{sys}$, in blue). We also overlay Kepler-1656c for comparison based on orbital properties derived in Section 2. We see from the leftmost panel in this figure that Kepler-1656b and Kepler-1656c must orbit with a nearly polar mutual inclination ($i_{bc} \approx 90°$) in order to be consistent with our observed $a_c = 3.05$ au and $e_c = 0.53$. We also see that for all panels, Kepler-1656c's orbital properties are consistent with models that reproduce Kepler-1656b's high eccentricity. Thus, EKL oscillations induced by gravitational perturbations from Kepler-1656c is a promising origin scenario for Kepler-1656b.

### 4.4. Predictions for Inner Planet Obliquity

One of the parameters of planet b that is traced in our simulations is the orbital obliquity, $\lambda_b$, which defines the angle between the host star's spin axis and the planet's orbital angular momentum vector. In our analysis, we found that the subset of simulations that are consistent with our observations produce a wide range of values for Kepler-1656b's current spin–orbit angle, or obliquity $\lambda_b$, with values ranging from prograde to retrograde and a slight preference for polar orbits ($\lambda_b \approx 90°$, see Figure 11). This, along with our predicted near-polar mutual inclination, is consistent with the previously suggested correlation between stellar obliquities and planetary mutual inclinations in systems with distant giant companions (Wang et al. 2022). Furthermore, a misaligned orbit for Kepler-1656b suggests that it is part of a growing sample of exoplanet systems hosting eccentric, misaligned, sub-Jovian planets with distant giant companions (see for example Yee et al. 2018; Dalal & Hébrard 2019; Correia et al. 2020). While it has been suggested that a slowly decaying outer protoplanetary disk may produce sub-Neptunes in these architectures, whether or not this mechanism is relevant for sub-Saturns like Kepler-1656b remains to be seen (e.g., Petrovich et al. 2020). Nevertheless, the misaligned and often polar orbits favored in our models show that gravitational interactions from a distant, outer companion may be the instigator for these configurations beyond the case of Kepler-1656b.

Figure 11 also demonstrates that engulfment and tidal locking produce distinctive obliquity distributions for the inner planet. For instance, systems for which the inner planet plunges into the host star are overwhelmingly on prograde orbits right before they plunge in. This trend should produce an observational signature on the population of post-engulfment host stars—specifically, plunging planets on prograde orbits will slow their host star's rotation (e.g., Qureshi et al. 2018; Stephan et al. 2020b). On the other hand, for systems in which the inner planet becomes tidally locked, the final distribution of obliquities matches that of tidally locked hot Jupiters (e.g., Naoz et al. 2012). These predicted obliquity trends will be relevant to the next-generation spectrometers that can resolve obliquities of Jovian and sub-Jovian planets in multiplanet systems.

## 5. Discussion

Our analysis suggests an in situ beginning (here we use "beginning" to refer specifically to the onset of EKL effects) with subsequent EKL-driven excitations is a plausible pathway for Kepler-1656b's orbital evolution to the close-in, highly eccentric orbit we see today. This proposed in situ configuration may point to in situ origins for Kepler-1656b, consistent with existing theories that dense planets with high core mass fractions can form in situ out of particularly dusty disk environments (e.g., Lee et al. 2014; Lee & Chiang 2015). It is also in line with more general theories of in situ giant formation (see, for example, Batygin et al. 2016). However, we cannot rule out the possibility that events such as flybys, scattering, or a merger occurred prior to EKL onset, in which case Kepler-1656b may not have formed in its present-day configuration. In fact, it is possible that a flyby or planet–planet scattering event pushed Kepler-1656b's orbit inwards from larger distances and/or into the required non-negligible mutual inclination between the two planets to trigger the onset of EKL (see Mustill et al. (2021), for an example of such evolution).[15] In this case, it remains possible that a merger prior to EKL onset explains Kepler-1656b's below-average envelope mass fraction ($f_{env} \approx 2.8\%$), as suggested by Millholland et al. (2020). Still, we stress that Kepler-1656b's eccentricity is not easily explained by flybys, scattering, or a merger alone—thus, perturbations from Kepler-1656c are our prevailing explanation for the inner orbital eccentricity.[16]

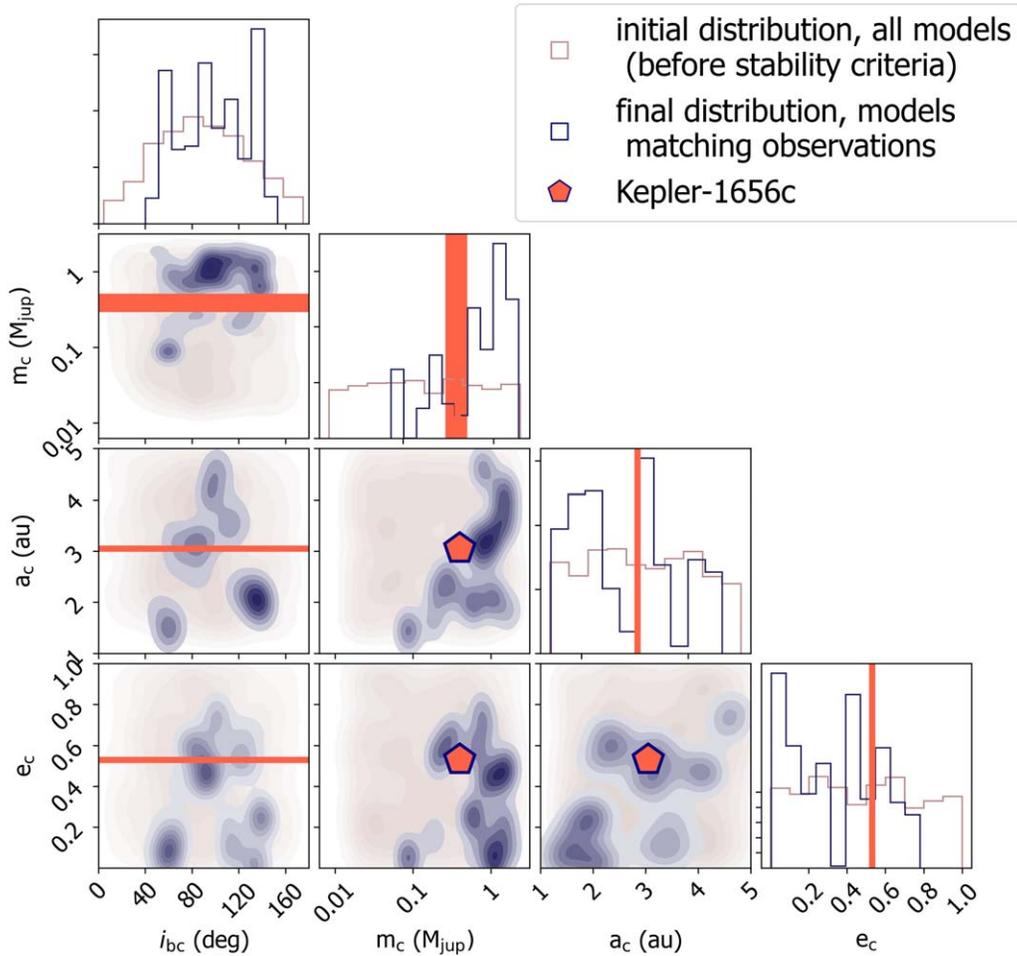

**Figure 10.** Smoothed 2D histograms of final conditions for all models (tan) and models that are consistent with our observations (blue) from Set 2 of our simulations are shown. We see that models consistent with our observations place constraints on the mutual inclination between Kepler-1656b and Kepler-1656c, which we refer to as $i_{bc}$. We also overlay Kepler-1656c based on properties found in Section 2. The diagonal panels show 1D distributions of parameters, that have been normalized to sum to 1.

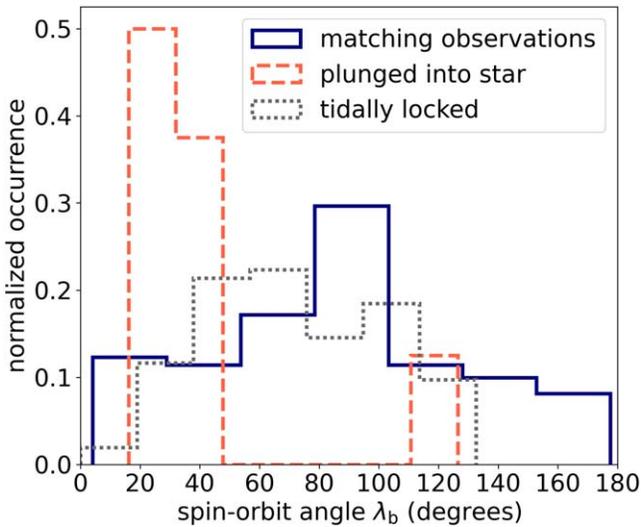

**Figure 11.** The final spin–orbit angle, or obliquity, of Kepler-1656b is shown for subsets of our simulations Set 2 for which the Kepler-1656b reproduces our observations (blue line), plunges into its host star (gray dotted line), or becomes tidally locked (red dashed line).

Previously, Brady et al. (2018) suggested that Kepler-1656b might be undergoing high-eccentricity migration given its proximity to the upper envelope of the e-a distribution. Here we find evidence for the contrary: our simulations indicate that Kepler-1656b's outer companion induces high-amplitude eccentricity oscillations gently, i.e., without triggering high-eccentricity migration. From these results, we classify two types of outer planets: those that induce tidal migration of their inner companions ("strong companions"), and those that do not, while still exciting eccentricity ("gentle companions"). We note that, due to the nature of their eccentricity oscillations, systems with gentle companions spend more time in less eccentric orbits, and only a small fraction of time in orbits as eccentric as Kepler-1656b's (see Section 3.1 and Figure 8). On the other hand, strong companions cause their inner planet to tidally circularize. Thus, our simulations show that sub-Saturns with outer companions can populate wide regions of *e-a* parameter space, depending on effects from the companion.

---

[15] The analysis of the possible flybys and/or planet–planet scattering prior to the EKL evolution is beyond the scope of this study.
[16] We note that stellar flybys are an unlikely scenario for our system as they needed an unreasonable number of close interactions (see Batygin et al. 2020).

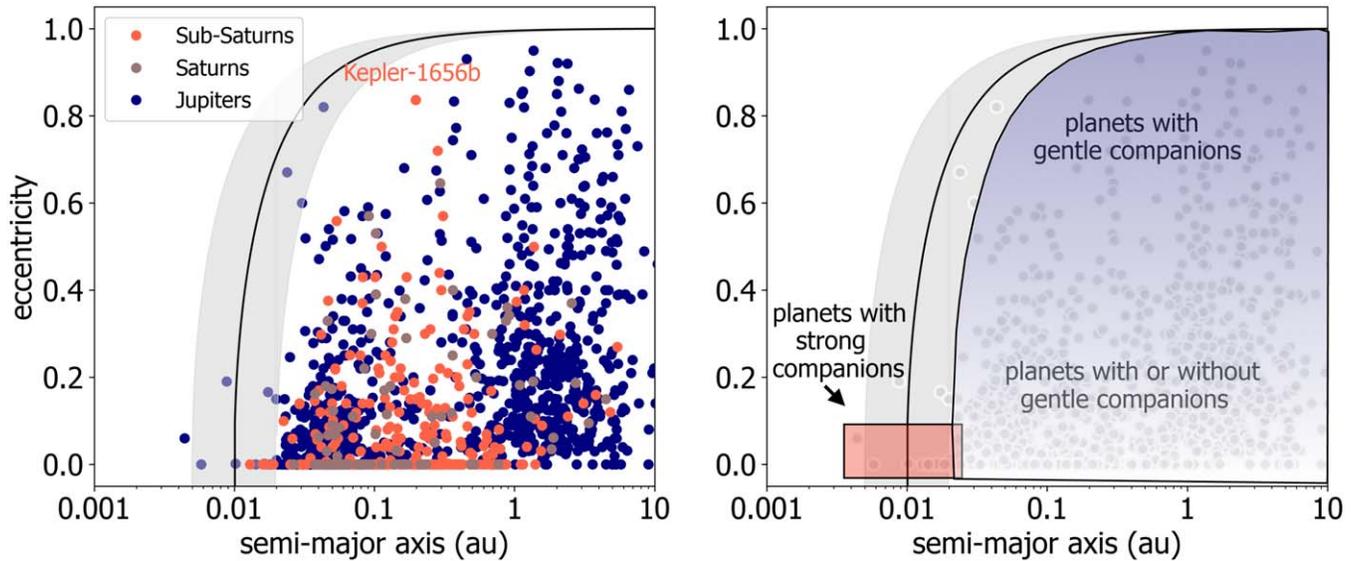

**Figure 12.** Left: the population of known giant planets is shown in *e-a* space. We plot subpopulations of sub-Saturns ($M = 15$–$60\ M_\oplus$) in blue, Saturns ($M = 60$–$100\ M_\oplus$) in red, and Jupiters ($M > 100\ M_\oplus$) in tan. Kepler-1656b and Kepler-1656c are overplotted as stars and colored according to mass. Data were obtained from the NASA Exoplanet Archive (Akeson et al. 2013). Right: the same population is plotted in gray. The shaded gradient represents an increasing need for gentle companions to explain orbital configurations toward the upper envelope, and the red region represents tidally locked configurations that may indicate the presence of strong companions. In both panels, the upper envelopes set by angular momentum conservation and chaotic tides are overplotted.

Comparing Kepler-1656b to the general population of sub-Saturns ($M = 15$–$60\ M_\oplus$), Saturns ($M = 60$–$100\ M_\oplus$), and Jupiters ($M > 100\ M_\oplus$) suggests similar dynamical histories for these planets, as highlighted in Figure 12. While a more rigorous statistical analysis is required, already from comparing our simulation results to observations (see Figure 12), the similarities in the *e-a* distribution of the sub-Saturns, Saturns, and Jupiters suggest that Saturn- and Jupiter-sized planets may be subject to similar perturbations from outer companions. Thus, a number of giant planets in the current exoplanet census may have orbital signatures of gentle and strong companions. In fact, a number of close-in, eccentric planets ranging from sub-Saturns to Jupiters have known companions that may be gently perturbing them into their eccentric configurations (e.g., CoRoT-20b, Kepler-419b, HD 156279b) or strongly driving them into a tidally locked orbit (see K2-43b, HAT-P-13b, Kepler-101b, HD 187123b). Beyond this, a larger subset of planets with eccentric or close, circular orbits do not have any confirmed companions (see for example GJ 2056 b, HD 20868b, HD 43197b, KOI-1257b, WASP-19b, TOI-1296b, TOI-1298b),[17] and would thus make a great laboratory to test our hypothesis of gentle and strong companions as the instigator for these configurations[18] Long-term monitoring of these planet systems' RV and transit data, combined with dynamical modeling of potential two-planet systems, can help us understand the origins of these systems and test our predicted positive correlation between eccentric or tidally sub-Saturns and outer companions.

Additionally, the three populations plotted in Figure 12 do not appear to show distinctive trends, although more data are needed to confirm this. Still, the apparent similarities in these populations of sub-Saturns, Saturns, and Jupiters in *e-a* space suggest that these planets are born from similar dynamical environments. This suggests that while these subpopulations of giant planets are often treated as distinct (e.g., Petigura et al. 2017a), they may be products of similar dynamical histories. Beyond this, it is thought that distant giant companions are responsible at least in part for the eccentricity distribution of small planets as well (Van Eylen et al. 2019). Future studies using a combination of observations and theoretical calculations to constrain outer companion properties will bring us closer to understanding the eccentricity distribution of giant planets and the exoplanet population as a whole, pulling the curtain back on the origins of these systems.

---

[17] Note that Moutou et al. (2021) implied that TOI-1296b and TOI-1298b's orbits may be connected to their high host star metallicity, implying that high metallicity produces more giant planet multiples (e.g., Dawson & Murray-Clay 2013). Still, based on the non-negligible fraction of locked planets in our simulations (10.3%), we suggest that TOI-1296b and TOI-1298b may have distant planetary companions.

[18] See the following references: CoRoT-20b, Deleuil et al. (2012); Kepler-419b, Dawson et al. (2014); HD 156279b, Díaz et al. (2012); K2-43b, Crossfield et al. (2016); HAT-P-13b, Bakos et al. (2009); Kepler-101b, Rowe et al. (2014); HD 187123b, Butler et al. (1998); GJ 2056b, Feng et al. (2020); HD 20868b, Moutou et al. (2009); HD 43179b, Naef et al. (2010); KOI-1257b, see Santerne et al. (2014); WASP-19b, Hebb et al. (2010); TOI-1296b and TOI-1298b, Moutou et al. (2021).

Our study demonstrates the advantages of combining observations and modeling to more rigidly constrain a planetary companion's origins and orbital properties (as highlighted in Figures 9 and 10). Here, we use observations to inform the types of outer planets in our simulations. From this, we uncover an in situ origin scenario for Kepler-1656b and make inferences about the inclination of the inner and outer planets in the system. Thus, our combined observational and theoretical analysis allows us to paint a detailed picture of the Kepler-1656 system's origins. These methods can be applied to similar single- and two-planet systems to constrain their architectures, histories, and properties of a potential companion.

I.A. is supported by the Dorothea Radcliffe Dea Fellowship and Dean's Scholarship at UCLA. We are grateful to the California Planet Search team and staff at W.M. Keck observatory for the remote observation efforts that made this study possible. I.A. would also like to thank Sanaea Rose, Judah Van Zandt, Dakotah Tyler, Jon Zink, Trevor David, and Sarah Millholland for helpful insights on the analysis and discussion sections of this paper. I.A. and S.N. acknowledge partial support from the NSF through grant No. AST-1739160. S.N. thanks Howard and Astrid Preston for their generous support.

*Software: Numpy* (van der Walt et al. 2011), *Radvel* (Fulton et al. 2018), *RVSearch* (Rosenthal et al. 2021), *corner* (Foreman-Mackey 2016), *astropy* (Astropy Collaboration et al. 2013, 2018).

## Appendix A
## Kepler-1656b Does Not Undergo High-eccentricity Migration

Under the physical conditions outlined in Section 3.1, Kepler-1656b's eccentricity can behave in several ways. In the quadrupole-level approximation, Kepler-1656b's eccentricity will oscillate between fixed minima and maxima. Alternatively, for conditions in which tides or octupole-level behavior are dominant, high-eccentricity migration or more chaotic eccentricity evolution can dominate. Consider the former case in which Kepler-1656b's eccentricity oscillates between a fixed minimum and maximum. In this scenario, Naoz et al. (2013a) showed that the total angular momentum is conserved, along with the eccentricity and angular momentum for the outer orbit. This implies that the initial and final mutual inclinations between the two planets (hereafter $i_{bc,i}$ and $i_{bc,f}$) obey

$$\cos(i_{bc,f}) = \frac{A_f \cos(i_{bc,i}) + L_1^2 e_1^2}{A_f \sqrt{1-e_b^2}}, \quad (A1)$$

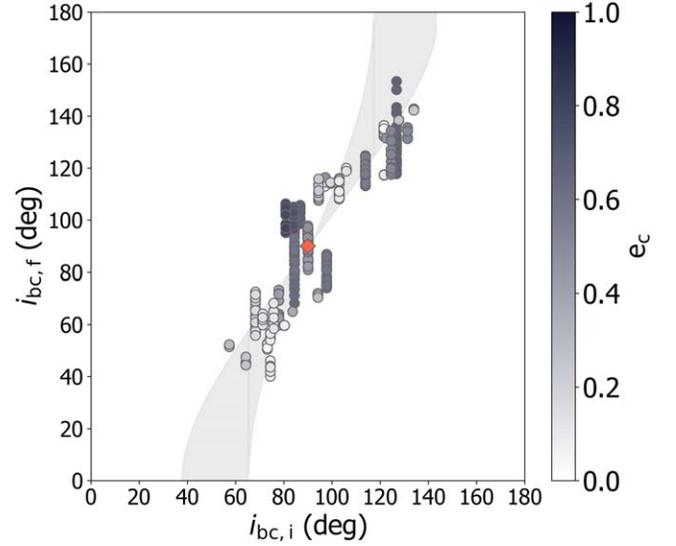

**Figure 13.** Initial and final mutual inclinations for all model timestamps consistent with observations of Kepler-1656b. The shaded region shows the initial and final inclinations predicted by the quadrupole-level approximation for inner planet eccentricities of $0.6 < e_b < 0.9$.

where $A_f = 2L_1 L_2^* \sqrt{1-e_c^2}$, and $L_1$ and $L_2$ are the respective conjugate angular momenta of the inner and outer orbit. Thus, if Kepler-1656b undergoes quadrupole-dominated eccentricity oscillations with no significant tides or octupole-level oscillations, it should obey the relationship described by Equation (A1). We plot this relationship for a range of eccentricities for Kepler-1656b that are reasonably close to our observed value ($0.6 < e_p < 0.9$) in Figure 13, and find that most simulations that match our observations lie in this regime. This indicates quadrupole-dominated behavior for Kepler-1656b, whereby the planet reached achieved its high eccentricity via sustained, high-amplitude eccentricity oscillations without tidal migration of chaotic behavior.

## Appendix B
## Correlated Parameters for Plunging, Locked Models

We find in our simulations that scenarios in which the inner planet plunges into the host star or becomes tidally locked arise from a unique set of parameters for the outer companion. This is evident in Figure 14, which shows the final distributions of companion parameters $e_c$, $m_c$, $a_c$, and $i_{bc}$ for both cases. We see here that moderately eccentric ($e_c \approx 0.5$), aligned companions are more likely to cause the inner planet to plunge into its host star. On the other hand, companions on polar orbits with a wider range of $e_c$ can cause the inner planet to become tidally locked. Thus, we predict that tidally locked sub-Saturns may be signatures of eccentric, polar companions.

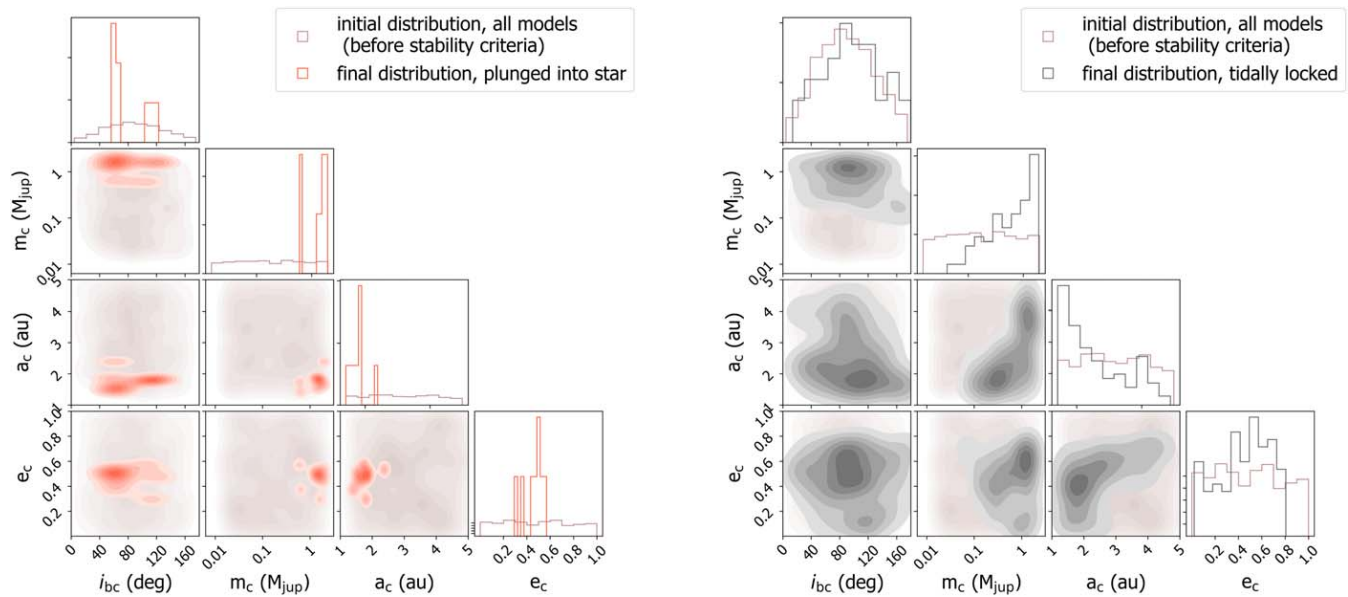

**Figure 14.** Smoothed 2D initial and final companion parameter distributions for models that plunged into their host star (left) and became tidally locked (right). The diagonal panels show 1D distributions of parameters, that have been normalized to sum to 1.

**ORCID iDs**

Isabel Angelo ⓘ https://orcid.org/0000-0002-9751-2664
Smadar Naoz ⓘ https://orcid.org/0000-0002-9802-9279
Erik Petigura ⓘ https://orcid.org/0000-0003-0967-2893
Mason MacDougall ⓘ https://orcid.org/0000-0003-2562-9043
Alexander P. Stephan ⓘ https://orcid.org/0000-0001-8220-0548
Howard Isaacson ⓘ https://orcid.org/0000-0002-0531-1073
Andrew W. Howard ⓘ https://orcid.org/0000-0001-8638-0320